\documentstyle[prl,aps,twocolumn,epsf]{revtex}

\begin{document}
\title{Coulomb blockade at a tunnel junction
between two quantum wires with long--range interaction}
\author{
Maura Sassetti\\
Istituto di Fisica di Ingegneria, INFM, CFSBT,
Universit\`a di Genova, Via Dodecaneso 33, I--16146 Genova, Italy}
\author{
Gianaurelio Cuniberti\\
Dipartimento di Fisica, INFM,
Universit\`a di Genova, Via Dodecaneso 33, I--16146 Genova, Italy}
\author{
Bernhard Kramer\\
I. Institut f\"ur Theoretische Physik,
Universit\"at Hamburg, Jungiusstra\ss{}e 9, D--20355 Hamburg, Germany}
\author{
{\small (Received 30 September 1996, to be published in Solid State 
Communications)
}
\vspace{3mm}
}
\author{
\parbox{14cm}{\parindent4mm\baselineskip11pt
%\begin{abstract}
{
\small The non--linear current--voltage characteristic of a tunnel junction 
between two  Luttinger systems is calculated for an interaction with
{\em finite range}. Coulomb blockade features are found. The
dissipative resistance, the capacitance and the external impedance,
which were  introduced {\em ad hoc} in earlier  theories, are obtained
in terms of the electron--electron interaction. The frequency
dependence of the impedance is given by the excitation spectrum of the
electrons. 
}
%\end{abstract}
\vspace{4mm}
}
}
\author{
\parbox{14cm}{
}
}
\maketitle

\vspace{1truecm}

The Coulomb blockade effect in the non-linear cur\-rent voltage ($I$-$U$)
characteristics of mesoscopic tunnel junctions 
\cite{likharev,mooijetal} has been the subject of many theoretical and
experimental investigations \cite{devoret,kramer} during the past decade.
Basically, due to the repulsion between the electrons,
tunneling is suppressed for voltages below $U_{C}= e/2C$, and 
temperatures smaller than
$T_{C}\equiv E_{C}/k_{B}$ ($k_{B}$ Boltzmann constant,
$e$ elementary charge, $C$ capacitance).
The quantity $E_{C}\equiv eU_{C}$ is the charging energy.

In the semi-phe\-no\-menological theory of the phe\-no\-me\-non \cite{ingold}
the tunnel junction is modelled by a capacitance and a tunnel resistance
$R_{t}$. An impedance $Z(\omega )$ is included
into the circuit. It represents the coupling of the tunneling
particles to a reservoir of Bosonic degrees of freedom. They guarantee
incoherence between different tunneling processes. When 
$Z(0)\equiv R=0$, the current voltage characteristic is linear,
$I(U)=U/R_{t}$. For $R\neq 0$, $I(U)\propto
U^{2R/R_{K}+1}$ when $U\ll U_{C}$ ($R_{K}$ von Klitzing constant), and
$I(U)\approx (U-U_{C})/R_{t}$ when $U_{C}\ll U\to \infty$. It must be
emphasized that the shift of the
linear behavior of $I(U)$ by $U_{C}$
is the important characteristic feature
of the Coulomb blockade for $R\to \infty$.

In this paper, we present a microscopic theory of the effect for a
one-dimensional (1D)  tunnel junction. The parameters introduced  in
the above mentioned theory by {\em ad hoc} assumptions are deduced 
consistently, and in a natural way, from the interaction between the 
electrons. 

Two semi-infinite (1D) systems of interacting electrons described  within
the Luttinger approximation  \cite{luttinger} are coupled by a tunnel
junction. The interaction potential between the
electrons is assumed to have a finite, non-zero range. The tunneling
current as a function of a voltage applied across the junction is
obtained.

The charging energy is
found to be the interaction potential
at zero distance, and the dissipative resistance is given by the spatial
average of the interaction potential. The spectrum of the elementary
excitations determines the impedance of the circuit. In order to
explain the latter no additional 'environmental modes' are needed.
The crucial point is that the above mentioned asymptotic behavior of
$I(U)$ for large $U$ appears to be directly related to the finite,
non-zero range of the interaction. The latter implies that the
dispersion relation of the elementary excitations of the electron
system becomes $\omega (k) \approx
v_{F}|k|$ in the short wavelength limit \cite{csk}. Our results show
that in 1D a charging energy, and, in turn, a capacitance can be
defined
%%%%
in the Luttinger model provided the interaction has 
%%%%
non-zero range.

We consider the Hamiltonian $H=H_{0} + H_{t} + H_{U}$. Here,
$H_{0}\equiv H_{el}^{(1)} + H_{el}^{(2)}$ consists of the Hamiltonians
of the two disconnected electron systems, which extend from $-L$ to $0$
and from $0$ to $L$ ($L\to \infty$), respectively. The tunnel junction
(at $x=0$) is described by $H_{t}$, and $H_{U}$ is the energy
contributed by the external voltage.

The electrons are described by the Bosonic
Luttinger Hamiltonian \cite{fabrizio},
($j=1,2$)
\begin{equation}
H_{el}^{(j)}=\sum _{q>0}\omega (q)\gamma ^{(j) \dagger}_{q}\gamma ^{(j)}_{q} +
\frac{\pi v_N}{4L}(\Delta N^{(j)})^{2} \;.
\end{equation} 
%%%%
Boundary conditions are assumed such that the original Fermion fields
vanish at $x=0,\pm L$. This introduces additional (quadratic)
off-diagonal terms in the Hamiltonian \cite{unpubl}. We neglect them
here for the sake of simplicity, since they do not affect the final
results qualitatively.
%%%%
The dispersion law 
$\omega (q)=q v(q)$ with the 
wave number dependent velocity
$v(q)=v_{F}[1 + \hat{V}(q)/\pi v_{F}]^{1/2}$,
reflects the 
Fourier transformed of the interaction potential $\hat{V}(q)$. 
For the latter, we assume a 3D screened Coulomb potential with the range $\alpha
^{-1}$ projected onto a quantum wire of diameter $d$.
%%%%%%
Depending on
whether $\alpha ^{-1}\ll d$ or $\alpha ^{-1}\gg d$ the interaction is
exponentially (Luttinger limit) or algebraically ($\propto x^{-1}$,
Coulomb limit)
decaying, respectively.
%%%%%%
For small
$q$ $(=n\pi/L$, $n$ integer) we
have the charge-sound excitations with the renormalized velocity
$v(0)\equiv v_F/g$
characteristic of the Luttinger system,
%%%%
with $g^{-1}\equiv [1 + \hat{V}(0)/\pi v_{F}]^{1/2}$. 
%%%%
For large $q$, we
find the excitation spectrum of the non-interacting electrons, due to
the finite range of the potential. The velocity associated to $\Delta N$
(see below) is $v_{N}=v(0)/g$.

The operators $\gamma ^{(j)\dagger}$, $\gamma ^{(j)}$
are related via a
Bogolubov transformation \cite{luttinger,fabrizio}
to the Fourier components of the phase fields,
$b^{(j)}_{q}\equiv \mbox{cosh}(\varphi _{q})\gamma ^{(j)}_{q}
- \mbox{sh}(\varphi _{q})\gamma ^{(j)\dagger}_{q}$, 
\begin{equation}
\Phi ^{(j)}(x)=\sum _{q>0}\sqrt{\frac{\pi }{qL}}\left(e^{iqx}b^{(j)}_{q}+
	e^{-iqx}b^{(j)\dagger}_{q}\right).
\end{equation}
They define (right moving) Fermion fields
\begin{equation}
\Psi ^{(j)}_{R}(x)=\frac{1}{\sqrt{2L}}e^{-i\vartheta ^{(j)}_{0}}
	e^{i\pi x\Delta N^{(j)}/L}
	e^{i\Phi ^{(j)}(x)},
\label{bosonized}
\end{equation}
with the density operators
$\rho ^{(j)}_{R}(x)=\Psi ^{(j)\dagger}_{R}(x)\Psi ^{(j)}_{R}(x)$.
The functions $\varphi _{q}$ contain the above dispersion relation,
$\exp{(-2\varphi _{q})}=\omega (q)/v_{F}q$.

The variables $\vartheta ^{(j)}_{0}$, defined modulo $2\pi $,
are conjugate to the number operators,
$\left[\vartheta ^{(j)}_{0}, \Delta N^{(j)}\right]=i$,
\begin{equation}
\Delta N^{(j)}\equiv 2L\left(\rho ^{(j)}_{R}-
	\frac{\partial_{x}\Phi ^{(j)}(x)}{2\pi}\right).
\label{chargedensity}
\end{equation}
The latter represent extra electrons in the
systems on the left and the right hand sides of the junction.

The above boundary conditions imply that
the corresponding left- and right moving parts are not independent but
$\Psi^{(j)}_{R}(x)=-\Psi ^{(j)}_{L}(-x)$, 
$\Psi ^{(j)}_{R}(x+2L)=\Psi ^{(j)}_{R}(x)$,
and either one of the two alone suffices to describe the system.
The tunnel Hamiltonian in terms of the latter is 
\cite{sassetti,fisher} 
\begin{equation}
H_{t}=L \Delta \left[\Psi ^{(2)\dagger}_{R}(0)\Psi ^{(1)}_{R}(0) +
\Psi ^{(1)\dagger}_{R}(0)\Psi ^{(2)}_{R}(0)\right]
\label{tunnel1}
\end{equation}
By inserting the above Bosonized form (\ref{bosonized}) one obtains
$
H_{t}\equiv H^{+}_{t}+H^{-}_{t},
$
with
\begin{equation}
H_{t}^{\pm}\equiv \frac{\Delta }{2}\exp{\left\{\pm i\sum_{j=1,2}(-1)^j
	[\vartheta ^{(j)}_{0}-\Phi ^{(j)}(0)]\right\}}.
\label{tunnel2}
\end{equation}

The electrostatic energy of the external voltage that is assumed to drop
only at the tunnel junction,
$U(x)=U\left[\Theta (x)-\Theta (-x)\right]/2$ ($\Theta (x)$
Heavyside function), is 
\begin{equation}
H_{U}\equiv -e\int_{-L}^{L}U(x)\rho (x),
\end{equation}
with $\rho (x)=\rho ^{(1)}(x)\Theta (-x) + \rho ^{(2)}(x)\Theta (x)$.
%%%%%%%
In
the dc limit, it has been shown that the current-voltage relation is
independent of {\em how} the voltage drops \cite{skprb,japanese}. Only
the voltage drop between $x\to -\infty$ and $x\to\infty$, which is
assumed to be fixed by an external ''battery'', is important.
Therefore, interaction induced rearrangement of charges in the presence
of the impurity is unimportant for the present calculation.
%%%%%%%
By inserting the above relation (\ref{chargedensity})
between $\rho^{(j)}_R$ and $\Delta N^{(j)}$, noting that
$\rho ^{(j)}(x)\equiv \rho ^{(j)}_{R}(x)+\rho ^{(j)}_{L}(x)=
\rho ^{(j)}_{R}(x)+\rho ^{(j)}_{R}(-x)$, and using 
$\Phi ^{(j)}_{R}(L)=\Phi ^{(j)}_{R}(-L)$ one obtains
\begin{equation}
H_{U}=\frac{eU}{2}\left( \Delta N^{(1)} - \Delta N^{(2)} \right).
\end{equation}

The current operator is defined as in previous works
$I\equiv ie[H^{-}_{t} - H^{+}_{t}]$ \cite{geigenmuller}.
Fermi's golden rule with
$H_{t}$ as a perturbation \cite{sassetti} yields the average current
\begin{equation}
I(U)=\frac{e\Delta ^{2}}{4}\left[1-e^{-\beta eU}\right]\int_{-\infty}^{\infty}
	dte^{ieUt}e^{-W_g(t)}
\label{IU}
\end{equation}
($\beta $ inverse temperature).
%%%%%%%
This equation is independent of the boundary conditions applied.
%%%%%%%
The thermal equilibrium correlation function 
\begin{equation}
W_{g}(t)\equiv \sum_{j=1,2}\left\langle\left[\Phi ^{(j)}(0)
-\Phi ^{(j)}(t)\right]\Phi ^{(j)}(0)\right\rangle
\end{equation}
is evaluated with respect to $H_0$.
The $\Phi ^{(j)}(t)$ evolve in the interaction picture 
with respect to $H_0+H_U$.
This gives
\begin{eqnarray}
&&W_{g}(t)=\int_{0}^{\infty} d\omega \frac{J(\omega )}{\omega ^{2}}
		\times\nonumber\\
&&\nonumber\\
&&\quad\times \left[(1-\cos{(\omega t)})
	\mbox{coth}\left(\frac{\beta \omega }{2}\right)
		+i\sin{(\omega t)}\right]
\label{correlation}
\end{eqnarray}

The key quantity is the spectral function $J(\omega )$ which 
is directly given in terms of the
dispersion law of the charge excitations of the system
\begin{equation}
J(\omega )=2\frac{\omega ^{3}q'(\omega )}{v_{F}q^{2}(\omega )},
\label{jofomega}
\end{equation}
where $q(\omega )$ is the inverse of the dispersion function $\omega
(q)$. Equation (\ref{jofomega}) establishes the main result of this
work in the sense that the quantity on the left hand side, which was
already present in the former theory \cite{ingold}, is now
directly related to spectrum of the elementary excitations of the
electron system \cite{csk}. 

\begin{figure}[htp]
%\vspace{5.5cm}
\label{spectral}
\centerline{
            \epsfxsize=8truecm
            \epsfbox{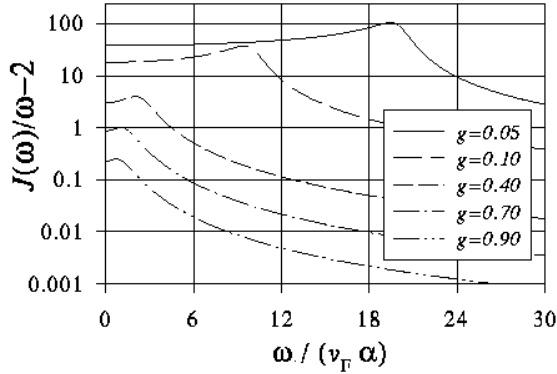}}
\caption[2]{The impedance function $J(\omega )/\omega -2$ of the
Luttinger limit 
for different interaction parameters $g$.
}
\end{figure}

The impedance function
$Z(\omega )\equiv J(\omega)/\omega-2$ for 
the interaction potential of the Luttinger
limit is shown in Figure \ref{spectral}.
Qualitatively similar curves are obtained for the Coulomb limit.
For $\omega \to 0$, $Z(0)=2(g^{-1}-1)\equiv R/R_{K}$. For large
$\omega $, $Z(\omega )$ tends to zero, i.~e. $J(\omega )\to 2\omega $,
the limit of non-interacting electrons. The peak in $Z(\omega )$,
which appears for smaller values of the interaction parameter $g$, is a
consequence of the inflection point in the spectrum of the elementary
excitations at $\omega _{p}$ (the frequency of the maximum in
$q'(\omega )$ \cite{csk}). When the screening
length decreases, $\omega _{p}$, and consequently the frequency beyond
which $Z(\omega )$ vanishes, increase proportional to $\alpha $
(Fig.~\ref{spectral}), for $\alpha \to \infty$, $Z(\omega )=2(g^{-1}-1)$.

In order to calculate the current voltage cha\-rac\-ter\-istic we have to
Fourier transform $\exp{[-W_{g}(t)]}$. This is done by using the relation
between $W_{1}(t)$, and the Fermi distribution $f(E)$\cite{footnote}, 
\begin{equation}
e^{-\frac{W_{1}(t)}{2}}=\frac{1}{\omega _{c}}
	\int_{-\infty}^{\infty}dE f(E)e^{iEt}.
\end{equation}
The result is
\begin{eqnarray}
&&I(U)=\frac{e\Delta ^{2}}{2\pi}
	(1-e^{-\beta eU})
\nonumber\\
&&\nonumber\\
&&\qquad\qquad\times\int_{-\infty}^{\infty} dEf(E)
	[1-f(E+eU)]
\nonumber\\
&&\nonumber\\
&&\qquad\qquad\times D(E)D(E+eU)
\end{eqnarray}
with the tunneling density of states
\begin{equation}
D(E)=\int_{-\infty}^{\infty}dt\cos{(Et)}e^{-W_{g}(t)/2}.
\end{equation} 
It is nothing but the local density of states at
the position of the tunnel junction \cite{sassetti,fisher,matveev}.

The connection with the earlier results 
\cite{ingold}
is established by observing that the
$I(U)$ can also be written as
\begin{eqnarray}
I(U)&=&\frac{1-e^{-\beta eU}}{eR_{t}}
	\int_{-\infty}^{\infty}dE\int_{-\infty}^{\infty}dE'
	f(E)[1-f(E')]\nonumber\\
&&\nonumber\\
&&\qquad\qquad\qquad\quad \times P(E+eU-E'),
\label{iofu}
\end{eqnarray}
with the probability density for a bulk excitation of energy $E$,
\begin{equation}
P(E)=\frac{1}{2\pi }\int_{-\infty}^{\infty}dt e^{iEt}e^{-[W_{g}(t)-W_{1}(t)]}.
\end{equation}
The tunnel resistance is $R_{t}\equiv 2\omega _{c}^{2}/e^{2}\Delta
^{2}\pi$ \cite{footnote}.

\begin{figure} [htp]
%\vspace{5.5cm}
\label{iu}
\centerline{
            \epsfxsize=8truecm
            \epsfbox{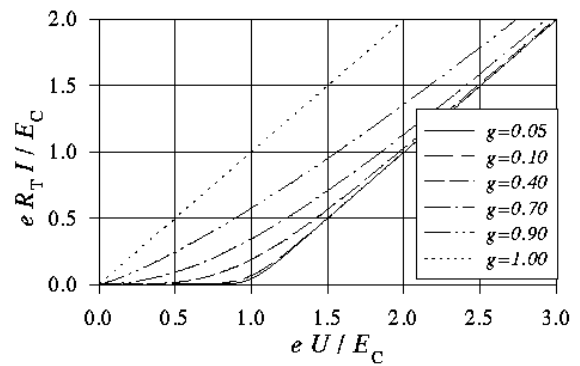}}
\caption[3]{The current voltage characteristic of a tunnel junction
between 1D interacting electron systems for different
strengths of the interaction, $g$.}
\end{figure}

In $P(E)$ 
the role of the 'electromagnetic environment' 
is now played by the
excitations of the interacting electrons. The result for zero temperature,
\begin{equation}
I(U)=\frac{1}{eR_{t}}\int_{0}^{eU}dE(eU -E)P(E),
\label{zerot}
\end{equation}
is shown in Fig.~\ref{iu} for different
interaction strengths.

For small voltages $I(U)\approx (1/R_{t})U^{2/g-1}$. By 
comparison with the corresponding
limit of reference \cite{ingold}, we recover the dissipative resistance
in terms of the interaction \cite{sassetti,matveev},
\begin{equation}
\frac{R}{2 R_{K}}\equiv \frac{1}{g} -1.
\label{resistance}
\end{equation}

For $U$ much larger than the range of $P(E)$ eq.~(\ref{zerot}) becomes
\begin{equation}
I(U)=
\frac{1}{R_{t}}\left(U-\frac{E_{C}}{e}\right),
\end{equation}
from which the charging energy is found 
by  using our microscopic expression (\ref{jofomega}) for
$J(\omega )$ 
\begin{equation}
E_{C}=\int_{0}^{\infty}d\omega Z(\omega )=2V(x=0).
\end{equation}
%%%%%%%
Taking into account the above mentioned off-diagonal terms induced by
the boundary conditions does not chan\-ge the dependence of $E_{C}$ on
the interaction potential. Only the prefactor is altered
\cite{unpubl}.
%%%%%

The capacitance found from $E_{C}$, $C=e^{2}/4V(0)$, scales in the same way as
the one, $C_{q}$, found previously by considering the frequency
dependent conductance of a pure quantum wire of interacting electrons
\cite{csk}. This was obtained by a completely different philosophy
starting from the classical theory of antennas. In fact, in the Coulomb
limit, $C/C_{q}=\mbox{O}(1)$. Following this approach,
one could consider the present problem as being the quantum equivalent
of two only very weakly coupled wires. From the comparison of their
AC properties with those of the quantum system considered here, one
would get the same scaling law for the capacitance.

The capacitance responsible for the Coulomb blockade is naturally given
here by the total capacitance of the circuit, i.~e. the two wires and
the junction. Apparently, in our ideal 1D model, the junction as such
does not contribute significantly. It is only needed
for detecting the Coulomb blockade induced by the latter. In the
Coulomb limit, $C=\pi \varepsilon \varepsilon _{0}d \approx 0.03$fF,
for a quantum wire fabricated in a semiconductor heterostructure,
AlGaAs/GaAs ($\varepsilon \approx 10$, $d\approx 100$nm), for instance,
with only one subband occupied. The charging energy is $E_{C}\approx
2.3$meV, which corresponds to a critical temperature of $T_{C}=30$K. By
decreasing the width of the wire, which seems achievable with present
days' technology, one should be able to increase $T_{C}$ close to room
temperature. Thus, a one-mode narrow quantum channel with 
a weak link appears to be the 'ultimate device' for observing Coulomb
blockade effects \cite{japanesepaper}.

From the AC properties \cite{csk}, we find also an inductance $L$ of
the electron system, due to the presence of the resonance in the
AC conductance at $\omega _{p}$ for strong interaction. It is
determined here by the ratio $E_{C}/\omega _{p}=e^{2}\sqrt{L/C}$, and is
reflected by the resonant behavior of the impedance $Z(\omega )$
(Fig.~\ref{spectral}). This is, however, of negligible
influence on the behavior of $I(U)$. Since
the integral weight of the resonance is small compared with the
charging energy, the total integral over $Z(\omega )$, and also
$\omega _{p}\ll E_{C}$ ($E_{C}\approx 500 \omega _{p}$ for the above
single mode wire), the steps in the derivative of the DC
current voltage characteristic predicted earlier as a signature of a
$\delta $-function like resonance in $Z(\omega )$ \cite{ingold} cannot
be obtained by using the present model. This prevents the inductance to
be detected directly here.

Our above results remain qualitatively true for a
non-linear dispersion relation, beyond  the Luttinger model, since the
influence of the interaction is negligibly small for wave numbers much
larger than the inverse of the range of the interaction. This is 
indicated by the large $q$ behavior of the dispersion with 
interaction in the random phase approximation \cite{sarma}.

For many channels occupied some of the abo\-ve con\-clu\-si\-ons
are mod\-i\-fied.
Mat\-ve\-ev and Glaz\-man \cite{matveev} trea\-ted the tun\-nel\-ing for a
qua\-si-1D quan\-tum wire with many ($N$) channels, but for zero range
interaction. They find a crossover in the zero frequency impedance
from eq.~(\ref{resistance}) to the non-interacting limit ($R\to 0$)
when $N\to \infty$. In that model, one cannot obtain the asymptotic
linear behavior of $I(U)$, even for finite $N$.
Therefore, we conclude that our above
result -- that using the Bosonization method the
Coulomb blockade is closely related
to the finite
range of the interaction -- remains also valid for many channels,
though the quantitative behavior of the charging energy, and thus the
capacitance, will be probably changed. We expect a contribution of the tunnel
junction to the capacitance proportional to the channel number. This
will be the subject of future studies \cite{unpubl}.

In summary, we obtained the non-linear current voltage characteristic
for a model of two 1D quantum wires of interacting electrons connected
via a tunnel junction. The features of the
Coulomb blockade phenomenon were found. The charging energy, the
capacitance, and the inductance of the circuit were given in terms of
the interaction potential.
%%%%%%%%
{\em  In order to obtain the Coulomb blockade in
the Luttinger liquid model, it is necessary to assume that
the interaction is of
finite, non-zero range}.
%%%%%%%
The role of the modes of the environment is
played by the elementary excitations of the interacting electrons.

This work has been supported by the SCIENCE and the HCM Programmes of the
EU, projects SCC--CT90--0020, CHRX--CT93--0136, repsectively,
and by the Deut\-sche For\-schungs\-ge\-mein\-schaft via the
Gra\-du\-ier\-ten\-kol\-leg ''Phy\-sik nano\-struk\-tu\-rier\-ter
Fest\-k\"or\-per'' of the Universit\"at Hamburg.

%%%%%%%%%%%%%%%%%%%%%%%%%%%%%%%%%%%%%%%%%%%%%%%%%%%%%%%%%%


\begin{thebibliography}{99}
\bibitem{footnote}Note that the usual high frequency cutoff $\omega
_{c} (\to \infty)$ is included here. It does not affect the
final results for the charging energy. It is only contained in
the tunnel resistance $R_{T}$ in the usual way
\protect\cite{ingold,sassetti}.
\end{thebibliography}
\end{document}